\documentclass[10pt,journal,compsoc]{IEEEtran}
\usepackage[T1]{fontenc}
\usepackage[utf8]{inputenc}
\usepackage{multirow}
\usepackage{graphicx}
\usepackage{amssymb}
\usepackage{xspace}
\usepackage[english]{babel}
\usepackage{flushend}
\usepackage[plain]{fancyref}
\usepackage{booktabs}
\usepackage{mdwlist}
\usepackage{tikz}
\usepackage{bibentry}
\usepackage{subfigure}
\usepackage{csquotes}
\usepackage{hyperref}

\newif\ifcomment%
\commenttrue%

\usepackage{colortbl}

\usepackage{wasysym}

\begin{document}
\definecolor{Gray}{gray}{0.9}
\pagenumbering{gobble}

\title{Moving Fast and Breaking Things: \\ How to stop crashing more than twice}

\author{Tobias~Fiebig \\ TU Delft \\
E-mail: t.fiebig@tudelft.nl
}
\maketitle

\begin{abstract}
``Moving fast, and breaking things'', instead of ``being safe and secure'', is \emph{the} credo of the IT industry.
In this paper, we take a look at how we keep falling for the same security issues, and what we can learn from aviation safety to learn building and operating IT systems securely.
We find that computer security should adopt the idea of safety.
This entails not only building systems that are operating as desired in the presence of an active attacker, but also building them in a way that they remain secure and operational in the presence of \emph{any} failure.
Furthermore, we propose a `clean slate policy design' to counter the current state of verbose, hardly followed best practices, together with an incident handling and reporting structure similar to that found in aviation safety.
\end{abstract}

\section{Introduction}
\label{sec:introduction}
In modern IT engineering, we like to believe that rapid development cycles are the only path to innovation, and readily accept the occasional failure while doing so~\cite{taplin2017move}.
In the minds of IT engineers and managers alike, this trade-off seems reasonable:
After all, what is the harm of occasional failure?
Our systems are dynamic, come with a full Continuous Integration/Continuous Delivery (CI/CD) pipeline, and a patch can be quickly deployed, making progress in the course of it.
Ultimately, it is not like lives depend on it, we tell ourselves.

However, the recent past has dramatically grounded this believe.
We saw people lose their freedom and live due to sloppy security choices in dating apps~\cite{grindr}.
We are still unable to grasp the true impact hundreds of millions of records exposed in data-breaches will have on the affected customers~\cite{hedley2017shape}.
We saw planes crash, due to errors and faulty trade-offs all too familiar to IT engineers. 

What all these incidents have in common is that they would have been preventable.
They are cases of operational misconfigurations, and simple misengineering that ignores best current practices.
Even worse, the mechanics underlying these incidents  are known for years and decades~\cite{dietrich2018investigating}.

To preserve and improve the security of information systems, various governance advances have been made to introduce regulation, processes and standardization.
Examples are ISO27001, \emph{Common Criteria} certification, and, most recently the European General Data Protection Regulation (GDPR).
Nevertheless, when we take a look at organizations security posture, we find that these regulations did not have the envisioned practical impact.
For examples, Equifax lost over hundred million individual data records due to a simple security blunder, while being ISO27001 certified~\cite{hedley2017shape}.
In fact, in a recent publication Dietrich et al.~\cite{dietrich2018investigating} find, that root-causes for security incidents in organizations are often not \emph{technical}, but in fact institutional issues.

In this article, we compare our observations on the current state of computer security with the early state of aviation safety, and find that they are comparable.
Based on this, we explore how and which governance levers could be adopted from the history of aviation safety to make IT systems more secure.

Our core-recommendations are:
\begin{itemize}
\item Including the notion of a system's safety into threat models.
\item Developing a common `clean-slate' best practices repository, by adding actionable practices based on what would have prevented investigated incidents.
\item Conducting public investigations of (major) IT security incidents, and feeding the results back to the best-practices repository.
\item Delegating regulatory and investigative authority for IT security incidents to a tree-formed structure of enforcement bodies, leveraging existing CSIRTs, under an international umbrella organization.
\item Reshaping the incentive-structure of computer security, by introducing fines that are initially low, but double for each repeated violation of guidelines from the global best-practices repository, to not inhibit innovation.
\end{itemize}

\noindent\textbf{Structure:} We discuss current examples of (repeated) computer security incidents in Section~\ref{sec:practice}, and compare to the development of aviation safety in Section~\ref{sec:planes}.
Based on this comparison, we assess practices from aviation safety and how these techniques may be beneficial to computer security in Section~\ref{sec:recommendations}.
Finally, we conclude in Section~\ref{sec:conclusion}.

\section{How Computer Security Fails}
\label{sec:practice}

The underlying premise of this paper is that computer security usually fails in \emph{preventable} ways. 
However, especially in \emph{hindsight}, everything can be considered preventable.
Hence, to define \emph{preventable} incidents, we first take a look at what we consider \emph{non-preventable} incidents.

In general, \emph{non-preventable} vulnerabilities include novel \emph{complex} attack opportunities, and new-found \emph{non-operational} vulnerabilities and exploits.
More to the point, every attack enabled by an 0-day, i.e., a vulnerability for which no mitigation is available, can be considered--at first--\emph{non-preventable}.
A good example for such an attack is the 2011 case of Northrop Grumman. 
In brief, state-level attackers utilized an 0-day in Adobe client software to breach the RSA Corporation, cloned an RSA 2-factor authentication token, and used that token to subsequently breach the defense contractor Northrop Grumman.
Of course, the attack could have been technically preventable, e.g., if clerk whose machine at RSA was initially infected had not opened the malicious attachment.
However, given the circumstances, no actor on the victims' side could have reasonably acted differently.
It is the clerks job to read external communication, and lateral movement in networks using further 0-days is hardly containable.

In contrast to this, \emph{preventable} security incidents are those where the attack could have been prevented all together, or the impact could have been significantly reduced by following best practices.
To underline our point that this definition fits far to many security issues of the recent past, we now take a brief look at three of them.
Specifically, we will take a look at the Equifax incident, Maersk being hit by NotPetya, and Grindr exposing millions of users to potential prosecution.


\subsection{Equifax Compromise: }
In mid-2017, Equifax---a US American credit rating agency---announced that unknown attackers stole personal data of over 143M American citizens and customers.
The stolen data contains critical personal information, including social security numbers, credit card numbers, living addresses etc. for nearly every person that conducted a credit based financial transaction in the united states during the last couple of years.
This includes, for example, non-residents that rented a hotel room using a credit card.~\cite{hedley2017shape}

\noindent\textbf{Why was it preventable?}
The attackers exploited a vulnerability in a component of the Apache Struts web-application server Equifax uses to run its main application.
However, a patch for this vulnerability was already available~\cite{hedley2017shape}, and should have been installed.
Yet, according to Equifax, a single employee forgot to take the necessary actions~\cite{hedley2017shape}.

The major issue here is that---as Equifax willfully acknowledges---an employee was able to \emph{forget} the update.
It is a well accepted best practice that the patch-level of IT systems should be monitored, and the responsible employee for upgrades notified by the monitoring system.
In case the employee does not react in time, the alert should be escalated to ensure that the (critical) update takes place.
This was obviously not done at Equifax, even though the same mechanic lead to countless incidents in the past~\cite{likeepers}.
Hence, we consider this incident to be \emph{preventable}.

\subsection{Maersk}
After the initial spread of 'NotPetya'---a potentially state sponsored ransomware from the 'Petya' family---the network of Maersk was compromised.
The infection vector was most likely MeDoc, a tax handling software used in Ukraine.
Via the Kiev office, NotPetya spread in the Maersk network, ultimately shutting down the company and causing between \$\(200M\) and \$\(300M\) in lost revenue.
Maersk's recovery efforts took weeks, with major parts of their central authentication system only being recoverable due to a single out-station being offline---and therefore not infected---due to a power failure~\cite{wired}. 

\noindent\textbf{Why was it preventable?}
The root cause of the initial infection was a compromised auto-update feature, i.e., a so called supply-chain-attack. 
In terms of defense this is close to impossible to mitigate.
However, the wider impact of this incident was enabled by insufficient network segmentation.
Following best practices for security of (critical) IT infrastructures, the Maersk network should have been \emph{at least} segmented and firewalled based on the national out-stations, ideally on a per-department level.
If this would have been done, the incident would have only affected Maersk Ukraine, and not the whole company.

Segmenting networks to stop the spread of worms (and other malicious activity) is a long standing best practice. 
As a community, we know this since the very first worm, the Morris worm, in 1988, and have---since then---reiterated this point tirelessly~\cite{orman2003morris}.
Yet, since then, countless companies fell victim to generations of worms due to insufficient network segmentation, with Maersk only being the newest victim in a long line of organizations suffering from the same issue.
Hence, again, we can consider this incident as \emph{preventable}.

\subsection{Grindr}
Grindr is a dating platform---while mostly being used by men---for queer people to find and contact potential partners in their vicinity.
Due to the wide-spread discrimination and criminalization of queer people around the world, it \emph{should} at least follow security best practices.
In fact, weak security of Grindr may actually threaten the lives of its users in several jurisdictions.
However, over the past years, Grindr continuously proved that they are rather following a na\"ive approach towards security.

The specific case we look at here is the security incident that forced Grindr to actually implement passwords: 
In 2012, they were using the International Mobile Equipment Identity (IMEI) of a users' phone for authentication and identification of close-by users.
In short, a user would send their IMEI to authenticate to the service, but also receive the IMEIs of other users in the vicinity along with their profiles.
While this was hidden in the interface, the IMEIs were the primary key for the App's user-list feature.
Made aware of this problem, Grindr opted to implement a self-rolled encryption layer to better hide this data from (malicious) users.
Unsurprisingly, two students from the University of Amsterdam were able to break this self-rolled encryption layer in 2013, and demonstrated how malicious users could take over Grindr users' accounts and read their private messages~\cite{grindr}.

\begin{figure}[t!]
\centering
\includegraphics[width=0.7\columnwidth]{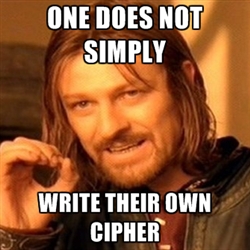}
\caption{A meme taken from \url{https://security.meta.stackexchange.com/a/915}, a stackexchange thread collecting in-community memes on bad security practices.}
\label{fig:comic}
\end{figure}

\noindent\textbf{Why was it preventable?}
In this case, Grindr failed twice.
First, they ignored the long standing practice of having at least 1-factor authentication, which sounds absurd in times of 2- and 3-factor authentication.
Instead, they used (semi) public information for authentication.
To mitigate this issue, Grindr opted to follow the two biggest fallacies in IT security:
Security by Obscurity in combination with self-rolled cryptography~\cite{grindr}.
Notably, especially the latter is consider an error so bold, that it made it into the programming community's strange humor, see Figure~\ref{fig:comic}.

\subsection{Skewed incentives and self-governance}
What all our examples have in common is that the incident, at least in its scale, would have been preventable if the affected company would have followed best practices.
Equifax should have not ``forgotten'' a security update, Maersk should have segmented their network, and Grindr should have used proper authentication methods from the beginning.
Yet, they did not follow best current practices for building and operating IT systems, and the question is \emph{`why'}?

To understand why best practices are often---seemingly---ignored, we have to get a high-level view on how risk management is usually done.
Risks are usually classified by their probability, i.e., how likely they occur, and their impact, i.e., how bad things get if the risk materializes.
Based on this classification, an organization can then decide what amount of resources it considers appropriate to mitigate the risk.

IT security has a problem on both sides of this equation. 
First, security, as practiced in IT operations, is often understood as an absence of errors.
This means that the return-on-investment is close to impossible to measure; in the best case, simply nothing happens.
However, this also means that the absence of security incidents due to random chance cannot be distinguished from an absence due to a securely build and operated system.
Second, the impact of security incidents is often negligible.
Lange \& Burger found a re-bound effect in stock prices of companies suffering from a security incident leaking user data.
In short, after an initial shock reaction, the companies value estimate will approach its valuation before the incident~\cite{lange2017long}.
Hence, it is naturally difficult to argue for any---usually costly---investment in IT security, given a low (perceived) event probability and a (perceived) unlikely loss due to the risk manifesting.


Looking at security incidents when end-users are affected, we find an interesting reasons for \emph{why} there was no large monetary impact, and therefore no incentive for change.
In case of Equifax, the simple reason is that no data of Equifax customers was lost.
Equifax' main revenue stream comes from credit rating services provided to banks, Mobile network operators, and other companies needing insights into end-users' credit score.
These companies are Equifax' direct customers.
However, the data Equifax lost is that of end-users.
Hence, end-users could only execute pressure on Equifax, i.e., vote with their feet, if they would (collectively) avoid signing contracts with companies using Equifax' services.
Given the power-imbalance in the market, this boils down to the question of `What do I value more: A new smartphone with my mobile contract, or my privacy?'
Considering users are willing to give up their passwords for some chocolate, this is mostly a rhetorical question~\cite{lesk2012price}.

If we extend this observation to the Grindr case, we find the stakes to be significantly higher.
Not only a user's privacy, but depending on the jurisdiction, their life is at stake.
Yet, this is just pulling the above example a little further. 
While in the previous example, the need was a simple function of social status, here we are talking about the opportunity of finding a partner.
In many, if not most, jurisdictions that discriminate against queer people, Grindr and similar apps may be the only way to do so.
So, like before, where the trade-off was `status vs. privacy', it is now `love vs. life.'

In contrast to the former two cases, Maersk obviously sticks out. 
While Grindr and Equifax clearly affected end-users, the biggest impact the Maersk incident had on their customers was that they were unable to bill them for the transport services they were providing.
In this case, the company suffering from the security incident also directly incurred a net loss.
However, we can assume that Maersk now takes following good security practices seriously. 
This is supported by research of Dietrich et al., who found that organizations suffering from a security incident will improve their security posture~\cite{dietrich2018investigating}.

So, in summary, we find that the incentive structure for IT security, especially when it comes to a large end-user base, is skewed. 
End-user behavior and post-incident damages do not necessarily tip the scales in terms of risk management.
Cases where there is a clear and measurable impact of security incidents in terms of revenue loss are surprisingly rare, but usually lead to an improvement in an organization's security posture~\cite{dietrich2018investigating}.
Hence, any approach to governing IT security should attempt to level the risk management field.
\section{A Short History of Aviation Safety}
\label{sec:planes}

In this section, we will explore how aviation safety developed over time---and how it was not so different than IT's current culture.

\subsection{Semantics of Safety}
When we talk about aviation \emph{safety} in comparison to computer \emph{security}, we must understand that `safe' and `secure' do not have the same meaning in aviation.
We---as IT engineers---usually think about security as the three basic properties we have to protect against an \emph{adversarial} attacker.
Yet, in aviation, safety means that a system continues to operate even in the presence of error, no matter if this error is intentional or not.
Their notion of `security' very much follows the same idea as ours, as IT engineers:
Systems have to be secure against tampering, by unauthorized or malicious personnel and external actors.
However, they also have to be safe.

\subsection{Early Aviation Safety}
The early history of aviation safety and security is surprisingly similar to the current state of computer security.
Specifically, aviation was an unregulated field of pioneers.
As such, safety was not a core concern. 
This changed with aviation becoming more and more prominent~\cite{heppenheimer1995turbulent}. 
By now, aviation safety is one of the most strongly regulated~\cite{amalberti1997human}, but also least error-prone fields of large system operations.
With aviation safety continuously improving over the last decades~\cite{amalberti1997human}, it is obvious that the mechanisms and policies to avoid preventable errors have been successful.

Following Heppenheimer, the notable governance actions that lead to a significant improvement in aviation safety too place between 1945 and 1965~\cite{heppenheimer1995turbulent}:
After the second world war, civil aviation for personal travel gained significant traction.
However, safety had not yet reached today's standards:
There were no procedures to analyze safety incidents, no rule-book that would ensure that errors only happen once.
As such, \emph{avoidable} incidents tended to reoccur~\cite{praetorius2011learning}.
In the next twenty years, several (global) policy decisions lead to the observed increase in safety, while the overall volume of aviation increased dramatically as well.

\subsection{Public post-mortems}
In the absence of pre-known best-practices to build on, aviation reached Today's safety levels mostly due to public and reactive post-mortems.
Reactive post-mortems, i.e., conducting a post-mortem to identify root-causes and then putting policies in place that prevent future similar incidents, are best illustrated by their history in aviation safety.
Heppenheimer reports on a case from 1960, where two planes collided over Brooklyn which mirrored the conditions of a similar incident four years earlier.
An investigation later concluded that the incident was avoidable~\cite{heppenheimer1995turbulent}.

However, as Heppenheimer notes, the most interesting thing about the 1960 incident was not what it changed, but what it re-affirmed:
That, in general, the incident was avoidable. 
However, this time around, the FAA (Federal Aviation Administration) conducted a thorough and \emph{public} root-cause analysis, deriving, mandating, and publishing \emph{rules} for those actions that would have prevented the incident~\cite{heppenheimer1995turbulent}.
The change to this methodology---deriving rules from a \emph{public} root-cause analysis and mandating their adoption---lead to a constant increase in aviation safety.
By 1989, aviation safety rose to a level where an avoided collision with two planes passing in a distance of two miles was considered as critical of an incident as the collision of two planes in 1960, triggering the same level of post-mortem analysis~\cite{heppenheimer1995turbulent}.

Note, however, that reactive post-mortems are detached from the actual mechanisms that have to be implemented.
While technical aspects, like check-lists, or standardization of procedures across actors, may also be adoptable to the computer security domain, we are focusing on the underlying governance approach:
Conducting a \emph{public} root-cause analysis, and mandating the adoption of preventive measures.

\subsection{Impact of Incidents}
On a first look, the severity of incidents is fundamentally different between computer security and aviation safety.
While faults in flight safety and security can lead to the death of passengers, issues in information security are often considered ``merely'' privacy issues---which they are not, as the Grindr case-study demonstrates.

However, not only since the IoT, safety and security in computer systems has been known to affect the lives of people.
One of the first, and probably most prominent computer safety incident leading to the death of people occurred in the early 1980's. 
A programming mistake in the control software of x-ray machines of a certain vendor lead to the death of several patients due to radiation over-exposure.
Today, computer systems control large (chemical) plants, medical systems, and other crucial aspects of human live.
For example, tampering with traffic-lights may lead to the road-death of countless people.
Hence, while it may seem like there is a huge difference in the possible impact of computer security and aviation safety, this is not the case.
Instead, the difference in potential impact is mostly a perception issue.


\subsection{Policy Enforcement}
In aviation safety, two major incentives/mechanisms that ensure the adoption of flight safety measures:
(i) The people affected by incidents (passengers) are also the direct customers of the airlines, and the threat of a plane falling to the ground is not as abstract as for computer security, see Section~\ref{sec:practice}. 
Hence, flight operators have a monetary incentive of increasing flight-safety.
(ii) There is an easy way of punishing policy violators by retracting landing rights.
In fact, one of the most effective means of enforcing aviation safety is a global blacklist of known best-practice violators.

In computer security, the impact of incidents is often abstract, and the affected users are not necessarily a companies' `customers.'
Furthermore, due to the global nature of the Internet, it is not as easy to effectively punish organizations that do not adopt best practices.
The FAA can prevent an airline with bad safety practices from landing at any US airport, thereby cutting them off the US market.
In comparison, it is hardly possible to prevent Europeans from accessing a social network service located outside of Europe if said service operator neglects computer security best practices.

\subsection{Public Authority}
With the International Civil Aviation Organization (ICAO) and International Air Transport Association (IATA), there are two international umbrella organization for flight safety and security with strong regulatory impact.
Commonly, there is a local government organization responsible for aviation safety, e.g., the FAA in the United States, and there is a clear chain of responsibility for investigating aviation safety incidents.

In contrast, there is no clear global and coordinated governance of computer security.
International standards organizations, like the IETF or ITU build Internet protocols, not focusing on concurrent events and incidents.
While---by now---local CSIRTs (For example the US-CERT in the United States) exist in almost all countries, they are not coupled with a strong reporting requirement for incidents involving actors from a specific country.
However, these reporting requirements would be necessary due to the reduced visibility of many computer security incidents in comparison with aviation safety incidents. 
Similarly, CSIRTs commonly do not hold investigative power to become active in case of an unreported incident.

Additionally, authority over incidents on the Internet cannot be easily defined. 
In aviation safety, the responsibility for investigating an incident lies with the country where the plane came down/where the incident happened. 
If this country is unable to perform a sufficient investigation, e.g., as in the case of MH17, this can be delegated to another country connected to the flight, i.e., the origin/destination country, or, the country the airline operating the plane has its seat in.

If we connect something to the Internet, it is basically everywhere.
In fact, we find that in areas where we do have some form of location based regulatory relevance, e.g., concerning data privacy laws or when it comes to taxing technology companies, actors try to ``select'' certain countries with a less stringent governance body or more lenient legislation.
For example, Facebook opted to have a European subsidiary in Ireland, so that the more relaxed data security regulations of Ireland are applicable for all European users.

\section{Lessons from Aviation}
\label{sec:recommendations}
So, what can we learn from aviation safety for IT security? 
How should we change and adjust our governance approach, and how can we make IT systems (more) safe?
We recommend five key items, regulators and engineers should consider to improve the global security of IT systems.

\subsection{Including safety in threat models} 
The first, relatively simple thing, that has to change is the role of error in our threat models. 
In fact, we claim that in addition to having threat models, we should start to develop safety models for IT systems and applications. 
These should, in absence of an attacker, assess how we risk failure in systems in the presence of (un)intentional errors, and---along the mitigations in threat-modelling---how we can mitigate these risks.
Hence, an error is not only `a bug' but an as serious incident as an attacker compromising a system in our world.
While changing this paradigm might not be hard on a \emph{technical} level---every IT engineer should already be able to build a safety model---it certainly requires significant effort in teaching, education, and awareness.

\subsection{Clean-slate Best Practices}
The field of IT is full of books, reports, blog-articles, and standards describing how things \emph{should} be done.
There are industry wide standards (ISO2700X, Common Criteria), and most recently, the GDPR \emph{should} force organizations to adopt best practices.
In practice, however, we find that even operators that are ISO/2700X certified, do still suffer from simple and avoidable mistakes.
For instance, Equifax, the main culprit in one of the most sever data leaks of the past years, claims to be ISO/27001 certified.

There seems to be something wrong with how we currently handle best practices.
However, looking at research by Dietrich et al.~\cite{dietrich2018investigating} on IT operations and earlier work by Hopkins on safety in hazardous industries~\cite{hopkins2011risk} shows that there can also be too many processes, procedures, standards, and best practices.
This also played a crucial role in the case of Equifax.
All the things they \emph{might} have done better aside, given their certifications it is not unlikely that they assumed they were already doing enough.
The crucial item they missed---a consistent and properly escalating monitoring system for security updates---might just have fallen through the cracks. 
As such, the case of Equifax is exemplary for the state in which we currently are in terms of IT operations and computer security:
There are too many, partly overlapping, partly disagreeing standards, certificates, procedures, and recommendations. 

Hence, we suggest to simply start from scratch. 
Instead of trying to condense the current abundance of best practices into a \emph{prescriptive} view on security, we create a \emph{descriptive} view on security by investigating incidents as they come along, mandating behavior that would have prevented the incident.
The specific form of the recommendations, i.e., whether it is adding checklists to procedures, mandating a strict four-eyes policy for critical procedures, etc.\@ should be considered based on the specific issues, and should be left open for later corrections.
One might argue, that our choice of a clean-slate for best-practices is somewhat na\"ive.
There already exists a plethora of recommendations, best practices, standards, and suggestions on how to properly run an IT system.
It would take too long to find a widely accepted common view on what and which of these should be included in an authoritative best-practices repository.
Hence, we claim that it will most likely be faster to just start from scratch.
After all, this is how aviation became safe, even though some of the lessons learned are---in retrospect---obvious, like `pilots should not be drunk.'

\subsection{Open Incident Investigations}
To enable our clean-slate best practices approach, we must be able to investigate and learn from incidents.
Hence, as in aviation safety, post-mortems should be open and public.

On its own, this recommendation sounds simple. 
Naturally, there already have been regulatory attempts towards this in the past.
Within the EU, several countries adopted policies that require companies to report security incidents, even before the GDPR.
Likewise, the DHS and US-CERT established a mandatory incident reporting procedure\footnote{\url{https://www.us-cert.gov/incident-notification-guidelines}}.
However, research by Laube \& B\"ohme shows that there are insufficient incentives for companies to participate~\cite{laube2016economics}, and, more importantly, these programs usually do not entail a \emph{public} database of incidents and recommendations.
Furthermore, as in other fields~\cite{flink2005lessons}, it is generally difficult what constitutes an incident worthy of reporting.
Not every SSH brute-force attack should go through a full incident review.

At the moment, some organizations, for example GitHub\footnote{See \url{https://github.blog/2018-10-30-oct21-post-incident-analysis/}.} already implement a public error culture. 
However, research by Dietrich et al. observes that many companies do not sufficiently budget for errors to enable blameless post-mortems~\cite{dietrich2018investigating}.
Hence, open incident investigations are not an isolated process.
They are a necessary steps for a clean-slate best practices repository, and they have to be implemented in conjunction with the right incentive models and proper delegation of regulatory control.

If we tie this back to the Maersk case from Section~\ref{sec:practice}, we see what public value could have been gained by a public incident investigation.
With its unsegmented network Maersk does not stand alone among Enterprises.
In addition, due to having to build their whole system from scratch, it is likely that Maersk now took a more segmented approach.
However, lessons learned from this process are not available to the wider public.
Naturally, one might argue that Maersk paid a lot for these insights, and therefore should retain them for themselves.
Yet, in aviation, letting everyone learn from a single actors mistake was crucial for developing today's safety levels. 

\subsection{Delegating Regulatory Authority}
Properly delegating the regulatory control is important when adopting a security and safety system modeled after the aviation industry.
In aviation, the ICAO (International Civil Aviation Organization) and the IATA (International Air Transport Association), are a public organization of aviation companies (IATA), and a United Nations backed governance body (ICAO).
Below that, we have the national organizations (like the FAA) and companies itself.
For the purpose of aviation, especially given their significantly lower frequency of incidents, this is a reasonable setup.

However, in computer security, we face a significantly higher number of incidents to be investigated, which means that our governance tree cannot be as flat.
In addition, there is not \emph{the} industry organization, or a comparable international governance body for computer security which could serve as an umbrella organization.
Hence, we face two distinct problems, or rather flows, of regulatory authority:
(i) The top-level governance, which concerns mostly building the best practices repository, and overseeing the global compliance of the lower levels, and,
(ii) The lower levels of incident investigation, and handling best-practices violations.
Furthermore, we have the issue of \emph{who} is responsible for handling an incident. 
As outlined in Section~\ref{sec:planes}, responsibility in aviation is usually rather clear:
The organization from the country where the incident happened, e.g., where the plane hit the ground, is responsible.

In general, we already have an active CSIRT community in most countries, which could pick up these tasks.
However, at the moment their key-focus is on dealing with incidents itself, after they occur, and providing support in adhering to existing security best practices.
Instead, to compare to aviation safety bodies, CSIRTs would have to have the mandate to:
(i) Investigating computer security incidents, with a legal backing to do a full root-cause analysis,
(ii) Identify actions that would have prevented the incident at hand, and communicate that upstream, and,
(iii) Being able to trigger sufficient action in case of best practice violations, i.e., fines.
Adjusting (national) CSIRTs rights and obligations, to leverage these existing structures would be essential in realizing our proposal.
Ultimately, the most important aspect is that this structure has to have the form of a tree, where investigations can be distributed, while results are communicated towards the root, and the root caters for the clean-slate best practices repository outlined above.
Hence, in summary, we suggest to delegate regulatory control \emph{bottom-up}, i.e., starting with \emph{national} regulations that delegate clear responsibilities for incident post-mortems to responsible and capable actors.
These then should also have the \emph{authority} and \emph{responsibility} to investigate IT security incidents in their jurisdiction, including applying fines.

At the same time, we have to build an \emph{international} organization as the root of this `tree' of CSIRTs, to serve as the ultimate `guardian' of the best practices repository.
The obvious solution to the first item would be reusing one of the international standards organizations---the ITU, the IETF, or maybe ISO---for this purpose. 
However, while it might not appear as easy, given these organizations track-record in finding consensus, it might be best to extend the clean-slate approach to this aspect, and create a dedicated body under the UN, as with ICAO.

At this stage, our proposal leaves the issue of who will ultimately be responsible for investigating an incident.
We suggest, that based on the severity of incidents, authority may remain in an organization, or the incident may have to be handled by a specific sector CSIRT.
The tree-shaped structure also allows us to create a responsibility structure that ensures global actors can no longer hide behind national governance inconsistencies.
In case of an incident spanning multiple countries, sectors, or organizations, the most covering node in the tree, becomes responsible for handling the incident.
For example, in case of a company active in several sectors within multiple countries in the European Union, ENISA, the European Union Agency for Cybersecurity, would be responsible. 
If an incident would involve countries outside of the European Union, or if the affected organization contests the investigation by ENISA, the international ICAO equivalent we suggest would have to take charge.
Similarly, an organization mainly active in the Netherlands' farming sector would be handled by the Dutch CSIRT for the producing sector, while they could ask for the case to be deferred to the Dutch NCSC.

\subsection{Creating Incentives for Secure Operations}
The most important governance change we suggest pertains incentives for following known best practices.
In Section~\ref{sec:practice}, we already discuss how---in computer security---the incentive structure for proper behavior is skewed. 
To address this issue, the GDPR already suggest fines of up to 4\% of an organization's annual revenue for serious data breaches.
However, we suggest to adjust this gradually, not only based on the severity and impact of an incident, but also based on whether it was preventable (as discussed in Section~\ref{sec:practice}) if the affected company would have followed the best-practices in-place at that time.
In fact, to facilitate a post-mortem culture driven by openness, and not fear, we suggest to \emph{not} fine organizations suffering from a data-breach due to an oversight not covered by codified best practices.
Instead, those first incidents should always lead to an update of these best practices.
Organizations, however, that do suffer from a data-breach due to violating best practices---maybe with an initial grace period after setting this policy in place---should be fined for, e.g., a minor fraction of their annual revenue.
This value should be doubled for any further incident due to a best-practices violation by the same organization.
The benefit of this slowly increasing fine-structure is that it neither punishes first-time offenders overly harsh, thereby not suffocating innovation, but at the same time puts a clear and measurable monetary incentive on repeated offenders, tipping the risk-equation in favor of computer security.

When we tie this back to the Grindr case discussed in Section~\ref{sec:practice}. 
Grindr, in itself, is an important tool within the Queer community.
This also explains why users are so unlikely to leave the service (see Section~\ref{sec:practice} for a discussion on the need-incentive issues users face.)
While their first error is \emph{bad}, immediately hitting the company with a 4\% annual revenue fine might have put them out-of-business.
At the same time, the actual turn of events---no fines at all---did not lead to them considering their computer security and users' privacy more seriously.
In fact, Grindr managed to become a press matter for security and privacy violations nearly every single year since 2012, with them, most recently, leaking their users' HIV status.
A progressively increasing fine-structure, in this case, would have ensured that the company gets a clear monetary incentive to behave securely, while not inhibiting the development of new and innovative platforms.

\section{Conclusion}
\label{sec:conclusion}

In this article, we suggest five mechanics available to policy makers to globally improve the state of computer security.
We suggest to adopt the safety-paradigm from aviation, along with their procedure of conducting public incident investigations, and creating codified best practices from these investigations.
However, we also find that two other aspects---regulatory control and inherent incentives for safe and secure behavior---are not directly adoptable. 
There are too many security incidents in computer security to adopt a similarly flat governance structure as in aviation, and the incentive system in terms of computer security is skewed towards inaction. 
In fact, it might just be cheaper for companies to not follow security best practices.
Hence, we suggest leveraging existing incident handling structures and communities in conjunction with a global umbrella organization modeled after the structure found in aviation to ensure global policy alignment.
This structure can then also be used to enforce globally comparable fine-based incentives for secure behavior. 

\emph{Next steps: }
In this paper, we present a vision of what \emph{might} improve IT security, based on the aviation sector's experiences. 
To realize our proposal in the future, we will have to conduct further research on the applicability of our recommendations.
Specifically, we must understand---experiences from other fields aside---how open incident investigations may improve IT security. 
Similarly, we have to understand how companies would react to the proposed fine based incentive model. 
However, as our recommendations are interdependent, the ultimate test of whether they work will be implementing them together.
We note that this will require a significant international collaboration effort. 
Hence, in the end, the question is `How much do we value secure and dependable IT systems?'

\section{Acknowledgements}
The ideas presented in this work have been discussed during Dagstuhl Seminar 19231.
I would like to thank the other participants for their insightful discussions and contributions to streamline my thoughts on this.
This work has been supported by the EU H2020 Safe-DEED project (Grant No. 830929), and the EU H2020 `CyberSecurity4Europe' pilot (Grant No. 825225).

\bibliographystyle{IEEEtranS}
\bibliography{paper,proceedings}

\vspace{0.001pt}
\end{document}